\documentstyle[twoside]{article}
\def\SU{{\rm SU}}

\newcommand{\md}[1]{\langle #1\rangle}
\makeatletter
\newcounter{alphaequation}[equation]
\def\thealphaequation{\theequation\alph{alphaequation}}
\def\eqnsystem#1{
\def\@eqnnum{{\rm (\thealphaequation)}}
\def\@@eqncr{\let\@tempa\relax
\ifcase\@eqcnt \def\@tempa{& & &}
\or \def\@tempa{& &}\or \def\@tempa{&}\fi\@tempa
\if@eqnsw\@eqnnum\refstepcounter{alphaequation}\fi
\global\@eqnswtrue\global\@eqcnt=0\cr}
\refstepcounter{equation}
\let\@currentlabel\theequation
\def\@tempb{#1}
\ifx\@tempb\empty\else\label{#1}\fi
\refstepcounter{alphaequation}
\let\@currentlabel\thealphaequation
\global\@eqnswtrue\global\@eqcnt=0
\tabskip\@centering\let\\=\@eqncr
$$\halign to \displaywidth\bgroup
  \@eqnsel\hskip\@centering
  $\displaystyle\tabskip\z@{##}$&\global\@eqcnt\@ne
  \hskip2\arraycolsep\hfil${##}$\hfil&
  \global\@eqcnt\tw@\hskip2\arraycolsep
  $\displaystyle\tabskip\z@{##}$\hfil
  \tabskip\@centering&\llap{##}\tabskip\z@\cr}

\def\endeqnsystem{\@@eqncr\egroup$$\global\@ignoretrue}
\makeatother

\begin{document}\large
\hfill\vbox{\baselineskip12pt
            \hbox{\bf IFUP -- TH. 7/94}
            \hbox{\bf INFN -- FE-01-94}
            \hbox{\bf LBL -- 35720}
            \hbox{\bf hep-ph/9405428}
            \hbox{May 1994}}
\vspace{7mm}
\begin{center}\vglue 0.6cm{\Large\bf\vglue 10pt
   Flavour in supersymmetric Grand Unification:\\ \vglue 3pt
   a democratic approach    \\}
\vglue 1.0cm
{\large\bf  Riccardo Barbieri, Gia Dvali$^*$, Alessandro Strumia,\\[4mm] }
\baselineskip=13pt {\em Dipartimento di Fisica, Universit\`a di Pisa
and \\[2mm] }
\baselineskip=12pt {\em INFN, Sezione di Pisa, I-56126 Pisa, Italy\\[6mm]}
\baselineskip=13pt {\large\bf Zurab Berezhiani$^*$,\\[4mm] }
\baselineskip=12pt {\em INFN, Sezione di Ferrara, I-44100 Ferrara,
Italy\\[4mm]}
\baselineskip=12pt {\large and\\[4mm] }
\baselineskip=13pt {\large\bf Lawrence Hall\\[4mm] }
\baselineskip=12pt {\em Department of Physics, University of California
at Berkeley, California 94720}

\vfill

{\large\bf Abstract}
\end{center}

\vglue 0.3cm{\rightskip=3pc \leftskip=3pc \tenrm\baselineskip=12pt
\noindent\large We consider the flavour problem in a supersymmetric Grand
Unified theory with gauged SU(6) group, where the Higgs doublets are understood
as pseudo-Goldstone bosons of a larger $\SU(6)\otimes\SU(6)$ global symmetry
of the Higgs superpotential.
A key element of this work is that we never appeal to any flavour symmetry.
One main interesting feature emerges: only one of the light fermions, an
up-type quark, to be identified with the top, can get a Yukawa
coupling at renormalizable level.
This fact, together with bottom-tau Yukawa unification, also implied in our
scheme, gives rise to a characteristic correlation between
the top and the Higgs mass.
By including a flavour-blind discrete symmetry and
requiring that all higher dimensional operators be mediated by the exchanges
of appropriate heavy multiplets, it is possible to give an approximate
description of all masses and mixing angles in term of a hierarchy of grand
unified scales. A special ``texture'' arises, implying a relation
between the top mass and the third generation mixing angles.
Several other possible consequences of this approach are pointed
out, concerning the $\mu/s$ mass ratio, the Cabibbo angle
and the proton decay.}

\vfill\normalsize\footnoterule~\\[0.5mm]
\noindent
{$\ast$~~Permanent address:\em{}
Institute of Physics, Georgian Academy of Sciences, 380077 Tbilisi, Georgia.}

\thispagestyle{empty}
\newpage
\setcounter{page}{1}

\normalsize
\section{Introduction}
It is definitely attractive to speculate that the physics of the elementary
particles, as seen nowadays in experiments, is only the extremely low energy
debris of a supersymmetric grand-unified world characterized by an energy
scale not far from the Planck mass itself. The experimental developments of
the last few years have brought some support to this view, more or less
directly. Although far from established, the perturbative nature of the
physical origin of the Fermi scale is more likely now than it was before the
electroweak precision tests performed especially at LEP~\cite{LEP}. On the
other hand, these same experiments, with the consequent highly precise
determination of the electroweak mixing angle, $\theta_{\rm W}$, confirm the
successful prediction of this quantity in supersymmetric
Grand-Unification~\cite{SUSYGUT}. Although the uncertainty of this
comparison, dominated by theoretical errors, has not been decreased
significantly, it has become nevertheless completely clear~\cite{GUT} that,
in the non supersymmetric case, it is possible to accommodate the measured
$\sin^2\theta_{\rm W}$ only at the price of introducing an intermediate
scale, namely one extra parameter. It is also clear how these
``signals'' for supersymmetry should turn into a direct experimental
evidence, in a positive or a negative sense: finding, or not finding, a light
Higgs and (some of) the superpartners at the Fermi scale.

{}From a theoretical point of  view, a weakness of supersymmetric Grand
Unification is the lack of clear progress in the understanding of flavour,
where most of the parameters of the current theory remain
undetermined\footnote{Bottom-tau Yukawa unification~\cite{b-tau} may
constitute a case of partial but significant progress.}.
The introduction, at the unification scale, of a large number
of Yukawa couplings with the same, or similar, unexplained patterns as
those ones of the Standard Model is an unsatisfactory feature of current
Grand Unified Theories. Even the appeal to their potential understanding in a
more fundamental underlying string theory may not satisfy many.

With these general motivations  in mind, we describe here an attempt to
improve on this situation. After a few considerations of general character,
we develop them in the case of an SU(6) Grand Unified theory, where the light
Higgs doublets are understood as pseudo-Goldstone bosons of a larger
$\SU(6)\otimes\SU(6)$ global symmetry of the Higgs
superpotential~\cite{SU6,SU6SU6}. The mechanism by which the Higgs doublets
are split from the unwanted colour triplets in this theory, (to be briefly
recalled later on), makes it appealing enough to be studied
further\footnote{The pseudo-Goldstone boson mechanism for the doublet-triplet
splitting in supersymmetric SU(5) Grand Unification
was first suggested in ref.~\cite{SU6gl}.
Our results, however, are specific of the gauged SU(6)
theory.}.

One immediate result of this approach is of great interest.
It is likely, in a sense that will be made
precise, that only one of the light fermions, an up-type quark, gets a Yukawa
coupling at renormalizable level: its identification with the top quark is
obvious. Crucial to this result is the fact that the top belongs to a
representation of the gauge group not isomorphic to those that
contain the other $Q=2/3$ quarks.
The smallness of the ratio between the bottom ($\lambda_b$) and the
top ($\lambda_t$) Yukawa coupling, together with bottom-tau Yukawa
unification, also implied in our scheme, requires, for consistency with the
observed bottom/tau mass ratio, that $\lambda_t$, naturally of order 1, be
actually at, or very close to, the infrared fixed point~\cite{IRfixed}.
In turn, this give rise to a correlation between the top and the
lightest Higgs mass, illustrated in figure~\ref{top-Higgs}.

{}From a general point of view,  our approach to the flavour problem can be
characterized as follows. An interesting attempt has been recently made by
Anderson et~al.~\cite{SO10} in supersymmetric SO(10).
They assume that a flavour symmetry will select very few~SO(10) operators,
perhaps the minimal number that can account for all masses and mixings of
quarks and leptons.
They find these operators to have a particular flavour dependence and a
particular dimensionality. It is an interesting problem to find out which
flavour symmetry can explain all these special features. To some extent, the
viewpoint adopted here is opposite to the one taken by Anderson et~al. We want
to find out, at least in the case of the SU(6) model, how far one can go in
understanding quark and lepton masses and mixings without ever introducing a
flavour symmetry. Is it possible to obtain a realistic theory of fermion
masses, with significant predictions, which results from writing the most
general Lagrangian involving a set of fields with given transformation under
the gauge group.

In section~2 we develop a few considerations relevant  for a generic
supersymmetric Grand Unified theory with heavy fermions and
non-renormalizable interactions scaled by inverse powers of the Planck mass.
In section~3 we focus on the SU(6) model, where we identify the light fields
and we study their Yukawa couplings induced by renormalizable interactions.
In section~4 we consider the possible effects of the non-renormalizable
interactions and we show how the ratio of the bottom, tau and charm masses to
the top mass can be related to the ratio of the SU(6) breaking scale to the
Planck mass.
In section~5 we study the possibility of generating the wanted operators
by means of appropriate heavy particle exchanges.
In section~6 we show how a specific ``texture'' for the fermion masses
arises.
Finally, in section~7, we study the couplings of the heavy coloured
triplet whose exchange is relevant to the proton decay amplitude
$p\to K\bar\nu$ and we identify a possible mechanism for its suppression,
intimately related with the origin of the doublet-triplet splitting.
Conclusions are given in section~8.

\begin{figure}\label{top-Higgs}
\caption{the correlation between the top and the lightest Higgs mass for
$\alpha_{\rm s}(M_Z)=0.110\div0.125$ (region between the dotted lines)
for heavy CP odd scalar.
The scalar partners of the top are taken unmixed and degenerate
with a mass of 1~TeV.
Also shown is the general upper bound on the Higgs mass in the MSSM.}
\end{figure}

\section{General considerations}
We consider in  this section a generic supersymmetric Grand-Unified theory,
based on a semisimple gauge group $G$ broken to
$\SU(3)\otimes\SU(2)\otimes{\rm U}(1)$ at the scale $M_G\approx 10^{16}$~GeV.
In general $G$ may undergo an intermediate step of breaking at a scale between
$M_G$ and $M_{\rm Pl}$. The successful unification of the SU(3), SU(2) and
U(1) gauge couplings constrains both the spectrum below $M_G$ and the group at
the intermediate stage of breaking. Below $M_G$ we consider the spectrum of
the Minimal Supersymmetric Standard Model, and we take the intermediate group
containing or coinciding with SU(5).

The vicinity of the Planck scale  to the unification mass suggests to
consider also the presence of possible non-renormalizable interactions scaled
by inverse powers of $M_{\rm Pl}$. Below the Planck scale, the theory in the
global supersymmetric limit will be characterized by the functions
$d(\hat z_i, \hat z_i^*)$ and $W(\hat z_i)$ of the chiral superfields
$\hat z_i$. (A third function, related to the kinetic term of the
gauge superfields, is irrelevant to the present discussion).
All fields with mass of order $M_{\rm Pl}$ have been integrated out and
their effects included in $d(\hat z_i, \hat z_i^*)$ and $W(\hat z_i)$. We
assume that the superfields can be redefined in such a way that
\begin{equation}\label{eq:d} d(\hat z_i, \hat z_i^*)=\sum_i\hat z_i\hat z_i^*
+ d^{\rm h.o.}(\hat z_i, \hat z_i^*) \end{equation}
where $d^{\rm h.o.}(\hat z_i, \hat z_i^*)$ only contains  higher order terms
in $1/M_{\rm Pl}$.

The chiral superfields $\hat z_i$ include the  ``matter superfields'' $\hat
f_a$ and the ``Higgs superfields'' $\hat H_n$.
The  scalar components of the Higgs superfields get vacuum expectation values
at the grand scale which respect supersymmetry and break $G$ to
$\SU(3)\otimes\SU(2)\otimes{\rm U}(1)$ with a possible intermediate step.
The $\hat H_n$ also contain the light SU(2) doublets $\hat h_1,\hat h_2$.
Under change of sign of the fields $\hat f_a$, which contain the  standard
quarks and leptons, the Lagrangian is taken to be invariant (``matter
parity''). We also assume that, in the supersymmetric limit, the Lagrangian
only contains the grand scale (with a possible fine structure) and that, in
the same limit, the quarks, the leptons and the $\hat h_1,\hat h_2$ doublet
superfields are exactly massless.

After insertion of the vacuum expectation  values of the Higgs fields at the
grand scale, $\md{H_n}$, the superpotential $W$ acquires the following
general form \begin{equation}\label{eq:WH}
W_{\md{H}}=\hat f_a M_{ab}\hat f_b + \hat f_a\lambda_{ab}^{(1)} \hat f_b\hat
h_1+ \hat f_a\lambda_{ab}^{(2)} \hat f_b\hat h_2+\cdots
\end{equation}
where $M,\lambda^{(1)},  \lambda^{(2)}$ are numerical matrices and all other
terms denoted by the dots (quartic terms in the matter fields, terms
involving the heavy Higgs fields, etc.) are irrelevant to the present
discussion.

By assumption, when reduced to diagonal form
\begin{equation}
M=U^* M_{\rm d} U^T,\qquad UU^\dagger=\hbox{1\kern-.24em I}
\end{equation}
the matrix $M$ has vanishing eigenvalues corresponding to
the standard quarks and leptons, $\hat f_\alpha$,
with the remaining eigenvalues, corresponding to heavy matter fields, of order
$M_G$.
By an obvious change of basis the superpotential in  the light fields reduces
to\footnote{The possible role of heavy fermion exchanges in the flavour
problem has already been explored in the literature~\cite{Exch}.}
\begin{equation}\label{eq:Wligh1}
W^{\rm light}=\hat f_\alpha(U^T \lambda^{(1)} U)_{\alpha\beta}
\hat f_\beta \hat h_1+
\hat f_\alpha(U^T \lambda^{(2)} U)_{\alpha\beta} \hat f_\beta \hat h_2
\end{equation}
In this superpotential, the heavy fields have been ``integrated out''.
It is at the same time remarkable  and non trivial that, if supersymmetry is
broken by a hidden supergravity sector~\cite{mM}, this same superpotential
with the only possible addition of a ``$\mu$-term'', $\mu \hat h_1 \hat h_2$,
(and $\mu$ of order of the effective supersymmetry breaking scale), describes
in the usual way also the supersymmetry breaking terms~\cite{SUGRA}.

For later purposes, it is useful to consider the particular case of a fine
structure at the grand scale, with an intermediate step of breaking induced
by the vacuum expectation values $\md{H_{n1}}\gg\md{H_{n2}}$.
Let us also suppose
that the mass matrix $M$ of the matter superfields has no other massless
eigenvector than the usual quarks and leptons, $\hat f_\alpha$, already at
the first step of breaking, when $\md{H_{n1}}\neq0$ but $\md{H_{n2}}=0$.
Among
the heavy fields, the only ones relevant to the present discussion are those,
$\hat F_a$, with the same $\SU(3)\otimes\SU(2)\otimes{\rm U}(1)$ quantum
numbers of standard quarks and leptons. In the superpotential with
$\md{H_{n1}}\neq0$, one has in general, in the appropriate basis
\begin{equation}
W_{\md{H_{n1}}}=\hat F_a^c M_{ab}\hat F_b + \hat
F_a\gamma_{a\alpha}^{(i)} \hat f_\alpha\hat h_i+ \hat
f_\alpha\lambda_{\alpha\beta}^{(i)} \hat f_\beta\hat h_i+\cdots
\end{equation}
where $\hat F_a^c$ are superfields transforming under
$\SU(3)\otimes\SU(2)\otimes{\rm U}(1)$ as the conjugate representation of
$\hat F_a$.
With also the smaller vacuum expectation values $\md{H_{n2}}$
different from zero, one has \begin{equation} W_{\md{H_{n1}},\md{H_{n2}}}=\hat
F^c (M+\delta M)\hat F +\hat F^c \mu \hat f +
 \hat F(\gamma^{(i)}+\delta \gamma^{(i)}) \hat f\hat h_i+
\hat f (\lambda^{(i)}+\delta \lambda^{(i)}) \hat f\hat h_i+\cdots
\end{equation}
where
$$\frac{\delta M}{M},\frac{\mu}{M},\frac{\delta\gamma}{\gamma},
\frac{\delta\lambda}{\lambda} \,\raisebox{-.4ex}{\rlap{$\sim$}}
\raisebox{.4ex}{$<$}\, {\cal O} (\frac{\md{H_{n2}}}{\md{H_{n1}}})$$
The mass term $\hat F^c \mu \hat f$ requires a redefinition of  the light
fields. To leading order in $\md{H_{n2}}/\md{H_{n1}}$ one has
\begin{equation}\label{eq:Wligh2}
W^{\rm light}=\hat f(\lambda^{(i)} +\delta\lambda^{(i)} -
\mu^T\frac{1}{M^T}\gamma^{(i)}) \hat f  \hat h_i. \end{equation}
Equations~(\ref{eq:Wligh1}) or~(\ref{eq:Wligh2}) give the effective  Yukawa
couplings of the light fields, a part from the need to bring to canonical form
the kinetic terms. In fact the higher order $d$-terms in eq.~(\ref{eq:d})
induce, as a consequence of the grand unified vacuum expectation values,
a wave function renormalization of the form $Z=1+\varepsilon$,  where
$\varepsilon$ is a hermitian matrix whose elements are at most of order
$\md{H}/M_{\rm Pl}$. To leading order in $\varepsilon$, it is easy to show
that the needed redefinition of the fields amounts to a shift of the light
masses by a factor $1+{\cal O}(\varepsilon)$ and to corrections of order
$\varepsilon$ of the Cabibbo-Kobayashi-Maskawa matrix.

\section{An SU(6) theory: renormalizable Yukawa couplings}
We want to consider the flavour problem in an SU(6) theory,  whose Higgs
system has been discussed in~\cite{SU6,SU6SU6}. The superpotential has
the form\footnote{The assumption of a superpotential which, in absence of
matter superfields, consists of a sum of two terms involving different sets
of superfields, including possibly some singlets, would be more attractive if
it could be shown to follow naturally from some symmetry of the theory.
Several possibilities can be envisaged, making use of continuous or discrete
invariances of normal or $R$-type nature.} \begin{equation}
W=W(\Sigma)+W(H,\bar H)  + W(\Sigma,H,\bar H, f_a) \end{equation}
where $\Sigma,H,\bar H$ are the Higgs superfields  transforming respectively
as 35, 6, $\bar 6$ representations of SU(6).
The part of the superpotential which does not depend on  the matter
superfields has a global $\SU(6)\otimes\SU(6)$ invariance.
Apart from $\SU(6)\otimes\SU(6)$ transformations,  the scalars in
$\Sigma,H,\bar H$ acquire the vacuum expectation values
\begin{equation}\label{eq:vevs}
\Sigma=\md{\Sigma}\,{\rm diag}\,(1,1,1,1,-2,-2),\qquad
H=\bar H = \md{H} \, (1,0,0,0,0,0).
\end{equation}
For the relative orientation between the $\Sigma$ and  the $H,\bar H$ vacuum
expectation values given in~(\ref{eq:vevs})\footnote{Such relative
orientation is fixed only after supersymmetry breaking by radiative
corrections. In a range of the low energy parameters, it has been shown that
such orientation, apart from corrections of order of the supersymmetry
breaking scale, corresponds to a local minimum of the effective
potential~\cite{SU6SU6}. The issue of the global minimum, which depends on
the heavy sector of the theory, will be discussed elsewhere.}, the gauge and
the global groups are broken as shown in table~1, where we have taken
$\md{H}>\md{\Sigma}$. (An opposite ordering of these vacuum expectation
values is not suggested by the unification of the coupling constants).
Of the Goldstone bosons
produced by the breaking of the global group, all are eaten but a pair of
SU(2) doublets, which play the role, after supersymmetry breaking, of the
light Higgs bosons. Their composition is (see also section~7)
\begin{equation} h_1=\frac{\md{H} h_\Sigma - 3\md{\Sigma} h_H}{\sqrt{\md{H}^2
+ 9\md{\Sigma}^2}},\qquad h_2=\frac{\md{H}\bar  h_\Sigma -3 \md{\Sigma}\bar
h_H}{\sqrt{\md{H}^2 + 9\md{\Sigma}^2}},\qquad \end{equation} in term of the
doublets (antidoublets) in $\Sigma$ and $H, \bar H$.

\begin{table}
\setlength{\unitlength}{1mm}\begin{center}\begin{picture}(0,0)
\put(22,-8){\vector(0,-1){12}}
\end{picture}\end{center}
$$\begin{array}{ccc}
\SU(6) & \SU(6)\otimes &   \hspace{-11.5ex}\SU(6)\\
\phantom{\md{H}}\downarrow\md{H} &\md{H}\downarrow\phantom{\md{H}} &
\hspace{-11.5ex}\\ \SU(5)&\SU(5)&
\hspace{-11.5ex}\phantom{\md{\Sigma}\downarrow}\md{\Sigma}\\
\phantom{\md{\Sigma}}\downarrow\md{\Sigma} &&   \hspace{-11.5ex} \\
\SU(3)\otimes\SU(2)\otimes{\rm U}(1) \qquad &&
\hspace{-11.5ex}\SU(4)\otimes\SU(2)\otimes{\rm U}(1) \end{array}
$$\caption{symmetry breaking pattern of the local and global groups.}
\end{table}

The chiral matter under SU(6) is normally contained  in $(15\oplus \bar 6
\oplus \bar 6')_i$ representations, where $i$ is a family index.
As it is well known, the representations $15\oplus \bar 6 \oplus \bar 6'$
 give the minimal anomaly free set
which has the standard 15-plet of chiral fields under
$\SU(3)\otimes\SU(2)\otimes{\rm U}(1)$.
We claim, however, that quarks and leptons do not necessarily live  in these
representations. This is because there are representations which, although
self-adjoint, do not contain the identity in their symmetric product and can
therefore obtain a heavy invariant mass only if they occur in even number.
The 20-dimensional tensor with 3~totally antisymmetric indices is the
smallest such representation in SU(6). We assume that they occur in odd
number. After integrating out all matter multiplets with Planck scale
invariant masses, we therefore consider\footnote{ We do not include any other
self-adjoint SU(6) representation, since the next one with no invariant mass
has dimension~540.} $$\hat f_a=(15\oplus \bar 6 \oplus \bar 6')_i \oplus
20\qquad i = 1,2,3$$ In a suitable basis, all possible SU(6) invariant Yukawa
couplings among these fields, assuming invariance under matter parity,
$\hat f_a\to-\hat f_a$, are
\begin{equation}\label{eq:W3}
W^{(3)}(\Sigma,H,\bar H,f_a)=\lambda^{(1)}
20~\Sigma~20+\lambda^{(2)}20~H~15_3+ \lambda^{(3)}_{ij}15_i~\bar{H}~\bar
6'_j. \end{equation}
For $\md{H}\gg\md{\Sigma}$, already at the first level
of breaking, from SU(6) to SU(5), the light quarks and leptons are identified
from this superpotential. Defining the SU(5) decomposition of the various
multiplets as \begin{equation}\begin{array}{lll}
20=10\oplus\overline{10} & 15_i=(10\oplus 5)_i\\
\bar{6}_i=(\bar 5\oplus 1)_i & \bar{6}'_i=(\bar 5\oplus 1)'_i \\
H = (5\oplus 1)_H & \bar H = (\bar{5}\oplus 1)_{\bar H} \\
\Sigma = (24\oplus 5\oplus \bar 5 \oplus 1)_\Sigma &
\end{array}\end{equation}
the mass term in eq.~(\ref{eq:WH}) has the form
\begin{equation}\label{eq:masses}
\hat f M\hat f = \lambda^{(2)} \md{H}\, 10_3\,\overline{10} +
 \lambda^{(3)}_{ij}\md{H}\, 5_i~\bar 5'_j,
\end{equation}
so that, apart from an irrelevant rotation in the heavy sector, $M$ is
already diagonal. The light quarks and leptons are in $10,10_1, 10_2$ and
$\bar 5_i$. Also light are all the SU(5) singlets $1_i $ and $1'_i$.
Accordingly, from eqs.~(\ref{eq:Wligh1}) and~(\ref{eq:W3}),  the only Yukawa
coupling between light states is contained in $\lambda^{(1)}20~\Sigma~20\to
\lambda^{(1)} 10~5_\Sigma~10$. One has therefore
\begin{equation}
W^{\rm light}=\lambda^{(1)}~Qu^c h_2
\end{equation}
where, under $\SU(3)\otimes\SU(2)\otimes{\rm U}(1)$,  $10=Q\oplus u^c\oplus
e^c$. At this stage only one up-type quark gets a mass after
$\SU(2)\otimes{\rm U}(1)$ breaking. This is an encouraging starting point:
with a Yukawa coupling of order~1, only the top gets a mass comparable with
the $W$-mass.
This important result depends crucially on two features of the theory.
First, the top quark must lie in an SU(6) representation which is not
isomorphic to those containing the charm and up quarks.
Second, one must take $\md{H}\gg\md{\Sigma}$.
If $\md{H}\ll\md{\Sigma}$ then none of the light quarks and leptons have
${\cal O}(1)$ Yukawa interactions with the Higgs doublet.
Also the case $\md{H}\ll\md{\Sigma}$ leads to an intermediate symmetry
group $\SU(4)\otimes\SU(2)\otimes{\rm U}(1)$, which does not
successfully predict the weak mixing angle.
Thus we discover that the requirement of a symmetry breaking
pattern which correctly predicts the weak mixing angle leads directly to
the prediction that just one of the quark and leptons has a mass
comparable to the weak scale, and this particle is a quark of charge 2/3.
We take $\md{H}\gg\md{\Sigma}$ for two reasons.
Firstly this leads to a reduction in the GUT scale threshold corrections
to the weak mixing angle.
Secondly, the two parameters $\varepsilon_H\equiv\md{H}/M_{\rm Pl}$ and
$\varepsilon_\Sigma\equiv\md{\Sigma}/M_{\rm Pl}$ will allow higher
dimension operators to account for the observed
hierarchy of quark and lepton masses, as will be shown in the next sections.


\section{Non renormalizable operators}
In fact,  non renormalizable operators scaled by inverse powers of $M_{\rm
Pl}$\footnote{We consider the Planck scale $M_{\rm Pl}$ as a natural scale
for these operators. One can bear in mind, however, that in certain cases the
regulator scale can be different (about $10^{18}\,{\rm GeV}$ in the context
of superstring inspired theories?).} give effects that cannot be ignored.
They can either be present in the basic theory, e.g.~as an effect of
gravitational interactions, or they can arise by integrating out heavy states
at the Planck mass. The following is a full list of all independent $f$-term
contributions involving four superfields
\begin{eqnarray}\label{eq:W4}
W^{(4)}(\Sigma,H,\bar H, f_a) &=&\nonumber
\frac{\gamma^{(1)}_{\alpha\beta}} {M_{\rm Pl}}
\bar{6}_\alpha~HH~\bar{6}_\beta + \frac{\gamma^{(2)}_i}{M_{\rm Pl}}
(20~\Sigma) H~15_i+ \frac{\gamma^{(3)}_i}{M_{\rm Pl}}20~H(\Sigma 15_i)+\\
&&+\frac{\gamma^{(4)}}{M_{\rm Pl}}20~H\bar H~20+
\frac{\gamma^{(5)}_{ij}}{M_{\rm Pl}} 15_i\bar H(\Sigma\bar{6}_j)+
\frac{\gamma^{(5')}_{ij}}{M_{\rm Pl}}15_i\bar H(\Sigma\bar{6}'_j)+\\
&&+
\frac{\gamma^{(6)}_{ij}}{M_{\rm Pl}}15_i (\Sigma\bar H)\bar{6}_j+
\frac{\gamma^{(6')}_{ij}}{M_{\rm Pl}}15_i (\Sigma\bar H)\bar{6}'_j\nonumber
\end{eqnarray}
where $\alpha,\beta=i,i'$.  In the ambiguous cases the parentheses denote, in
a self-explanatory notation, how the SU(6) indices are contracted.
In next order,  the terms giving the dominant contributions to the masses of
the light states are \begin{eqnarray}\label{eq:W5}
W^{(5)}(\Sigma,H,\bar H, f_a) &=&\nonumber
\frac{\sigma^{(1)}_{ij}}{M_{\rm Pl}^2}15_i~H\Sigma H~15_j +
\frac{\sigma^{(2)}_i}{M_{\rm Pl}^2}20~\bar H \Sigma\bar H\bar{6}_i+\\
&&+\frac{\sigma^{(3)}_{ij}}{M_{\rm Pl}^2}15_i~(\Sigma^2 \bar H)~\bar{6}_j +
\frac{\sigma^{(4)}_{ij}}{M_{\rm Pl}^2}15_i~(\Sigma\bar H)(\Sigma\bar{6}_j)
\end{eqnarray}
If all these operators are present,  as indicated, with dimensionless
couplings of order unity and no particular flavour structure, one does not
get the correct pattern of masses for two reasons. The operator
$15(\Sigma\bar{H})\bar{6}$ would give the dominant mass to down-type quarks
not sitting in the same SU(2) left-handed multiplet as the top. Furthermore,
all operators depending on 2 indices $i,j$ would give comparable masses to
all generations.
We therefore conclude that a completely general operator analysis
based on this set of fields lighter than the Planck scale cannot explain
the observed fermion masses.
The next step is to study whether a restricted set of operators can yield
acceptable masses.
If so the origin of this restricted set will have to be addressed.

Consider the theory in which $\gamma^{(5)},\gamma^{(6)},\sigma^{(1)},
\sigma^{(3)}$ and $\sigma^{(4)}$ are absent.
In this case only $\gamma^{(2)}$ and $\sigma^{(2)}$ lead to important
mass terms. Taking into account that the multiplet $15_3$
has already been defined by $W^{(3)}$, eq.~(\ref{eq:W3}),
without loss of generality, we can redefine
\begin{equation} \begin{array}{l}\displaystyle
\frac{\gamma^{(2)}_i}{M_{\rm Pl}} (20~\Sigma) H~15_i\to
\frac{\gamma^{(2)}_3}{M_{\rm Pl}} (20~\Sigma) H~15_3+
\frac{\gamma^{(2)}_2}{M_{\rm Pl}} (20~\Sigma) H~15_2\\  \displaystyle
\frac{\sigma^{(2)}_i}{M_{\rm Pl}^2}20~\bar H \Sigma\bar H\bar{6}_i\to
\frac{\sigma^{(2)}}{M_{\rm Pl}^2}20~\bar H \Sigma\bar H\bar{6}_3
\end{array}
\end{equation}
In this way we obtain a  massless first generation of quarks and leptons and,
using eq.~(\ref{eq:Wligh2}),  the following Yukawa couplings for the $2^{\rm
nd}$ and $3^{\rm rd}$ generations\footnote{The combinatoric factors in front
of the various operators, as well as of the kinetic terms, not shown, are the
inverse of the number of all possible irrelevant permutations of fields and
SU(6) indices.} ($\varepsilon_H\equiv\md{H}/M_{\rm Pl}$)
\begin{eqnarray}\nonumber W^{\rm light} &=& \left[
\lambda^{(1)}\frac{1}{2\cdot 2} 20^{abc}\Sigma_c^d 20_{dab} +
\frac{\gamma_2^{(2)}}{M_{\rm Pl}}\frac{1}{2} 20^{abc} \Sigma_c^d H_d
15_{bc}^2 + \frac{\sigma^{(2)} }{M_{\rm Pl}^2} \bar H^i 20_{ijk} \Sigma^j_l
\bar H^l \bar 6^k_3\right]_{\rm light} \\
 &=& (\lambda^{(1)} Q u^c + \gamma_2^{(2)} \varepsilon_H Q u_2^c +
\label{eq:Flavour}
\gamma_2^{(2)}\varepsilon_H Q_2 u^c) h_2+\sigma^{(2)}\varepsilon_H^2
(Qd_3^c + e_c L_3) h_1
\end{eqnarray}
For the decomposition under  $\SU(3)\otimes\SU(2)\otimes{\rm U}(1)$ we have
defined \begin{equation}
\begin{array}{l}
10_i = (Q+u^c + e^c)_i\\
\phantom{1}\bar{5}_i = (d^c + L)_i
\end{array}
\end{equation}
in analogy with (for the 10 in the 20)
\begin{equation}
10=Q+u^c+ e^c
\end{equation}
Let us now recall, from  standard renormalization group
(RG) results~\cite{RGE}, the connection between the running  masses,
$m_t,m_c,m_b,m_\tau$, and the corresponding Yukawa couplings at the
unification scale ($\tan\beta\equiv\md{h_2}/\md{h_1}$): \begin{eqnarray*}
m_t  &=&  v\sin\beta \lambda_t \zeta^u y^6\eta_t \\
m_c = 1.27\pm0.05\,{\rm GeV}  &=&  v\sin\beta \lambda_c\zeta^u y^3\eta_c\\
m_b = 4.25\pm0.10\,{\rm GeV} &=&   v\cos\beta \lambda_b\zeta^d y\phantom{^2}
\eta_b \\ m_\tau = 1.78 \,{\rm GeV}  &=&
v\cos\beta\lambda_\tau\zeta^e\phantom{y^2}\eta_\tau\\ V_{cb}=0.04\pm0.01 &=&
V_{cb}(M_G)/y \end{eqnarray*}
where
$$y\equiv\exp\left[-\int_{\ln M_Z}^{\ln M_G}
\frac{\lambda_t^2(\mu)}{16\pi^2} d\ln\mu\right]$$ and, for $\alpha_{\rm
s}=0.110\div 0.125$, $$\eta_t\approx 1,\qquad
\eta_c=1.87\div2.30,\qquad
\eta_b=1.47\div 1.62,\qquad
\eta_\tau=0.991,$$
$$\zeta^u=3.4\div3.8,\qquad \zeta^d=3.3\div3.7,\qquad \zeta^e=1.5.$$
We therefore obtain, on the basis of equation~(\ref{eq:Flavour}):
\begin{itemize}
\item[i)] from $m_b/m_\tau$: $\lambda_t=\lambda^{(1)}=2\div5$;
\item[ii)] from $m_c/m_b$: $\tan\beta\lambda_c/\lambda_b=
\tan\beta (\gamma_2^{(2)})^2/(\lambda^{(1)}\sigma^{(2)})=0.55\pm0.20$;
\item[iii)] from $m_\tau$:
\begin{equation}\label{eq:eH}
\lambda_\tau \cos\beta=\varepsilon_H^2 \sigma^{(2)} \cos\beta=0.0070,\qquad
\varepsilon_H\ge\frac{0.083}{\sqrt{\sigma^{(2)}}};
\end{equation}
\item[iv)] from $|V_{cb}|$:
$\varepsilon_H  \gamma_2^{(2)} /\lambda^{(1)} =0.027\pm0.010$.
\end{itemize}
This gives an elegant understanding of all heavy fermion masses in terms of
the ratio $\varepsilon_H=\md{H}/M_{\rm Pl}$ and of parameters of order unity.
As it is well known~\cite{IRfixed}, bottom-tau Yukawa unification and moderate
$\tan\beta$, both implied in our scheme, require $\lambda^{(1)}$ not only to
be of order~1, but actually close to its infrared fixed point, and,
consequently, for the physical top mass
\begin{equation}\label{eq:mt}
M_t=m_t
[1+\frac{4}{3\pi}\alpha_3(m_t)]=
\sin\beta\,(190\div 205)\,{\rm GeV}=140\div205\,{\rm GeV}
\end{equation}
Even more interesting is the correlation between
the masses of the top and of the lightest Higgs boson.
Since the top Yukawa coupling at the low scale is essentially fixed, such
correlation is only determined by the value of $\tan\beta$ and is shown in
figure~\ref{top-Higgs} for $\alpha_{\rm s}(M_Z)=0.110\div0.125$.
As a matter of fact, what is shown in figure~\ref{top-Higgs}
is an upper bound on the lightest scalar Higgs mass, which can be lowered by
having an equally light pseudoscalar in a well known way.
The scalar partners of the top are taken unmixed and degenerate
with a mass of 1~TeV~\cite{OneLoop}.

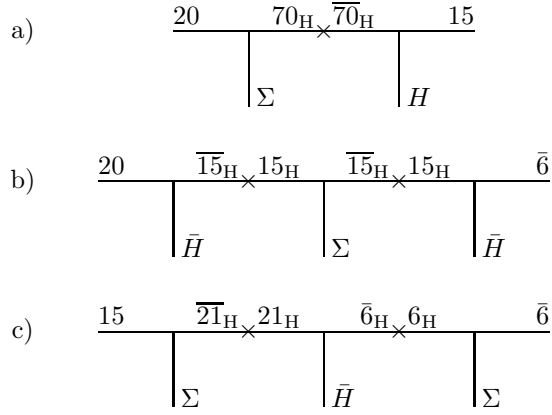
\begin{figure}\setlength{\unitlength}{1cm}
\begin{center}\begin{picture}(6,5)(0,-5)
\put(-1,-0){\makebox(0,0){a)}}
\put(-1,-2){\makebox(0,0){b)}}
\put(-1,-4){\makebox(0,0){c)}}
\put(1,0){\line(1,0){4}}
\multiput(2,0)(2,0){2}{\line(0,-1){1}}
\put(1,0.1){\makebox(0,0)[bl]{$20$}}
\put(3,0.1){\makebox(0,0)[b]{$70_{\rm H}~~\overline{70}_{\rm H}$}}
\put(5,0.1){\makebox(0,0)[br]{$15$}}
\put(2.1,-1){\makebox(0,0)[bl]{$\Sigma$}}
\put(4.1,-1){\makebox(0,0)[bl]{$H$}}
\put(3,0){\makebox(0,0){$\times$}}
\put(0,-2){\line(1,0){6}}
\multiput(1,-2)(2,0){3}{\line(0,-1){1}}
\put(0,-1.9){\makebox(0,0)[bl]{20}}
\put(2,-1.9){\makebox(0,0)[b]{$\overline{15}_{\rm H}~~15_{\rm H}$}}
\put(4,-1.9){\makebox(0,0)[b]{$\overline{15}_{\rm H}~~15_{\rm H}$}}
\put(6,-1.9){\makebox(0,0)[br]{$\bar{6}$}}
\put(1.1,-3){\makebox(0,0)[bl]{$\bar{H}$}}
\put(3.1,-3){\makebox(0,0)[bl]{$\Sigma$}}
\put(5.1,-3){\makebox(0,0)[bl]{$\bar{H}$}}
\put(2,-2){\makebox(0,0){$\times$}}
\put(4,-2){\makebox(0,0){$\times$}}
\put(0,-4){\line(1,0){6}}
\multiput(1,-4)(2,0){3}{\line(0,-1){1}}
\put(0,-3.9){\makebox(0,0)[bl]{15}}
\put(2,-3.9){\makebox(0,0)[b]{$\overline{21}_{\rm H}~~21_{\rm H}$}}
\put(4,-3.9){\makebox(0,0)[b]{$\bar{6}_{\rm H}~~6_{\rm H}$}}
\put(6,-3.9){\makebox(0,0)[br]{$\bar{6}$}}
\put(1.1,-5){\makebox(0,0)[bl]{$\Sigma$}}
\put(3.1,-5){\makebox(0,0)[bl]{$\bar{H}$}}
\put(5.1,-5){\makebox(0,0)[bl]{$\Sigma$}}
\put(2,-4){\makebox(0,0){$\times$}}
\put(4,-4){\makebox(0,0){$\times$}}
\end{picture}
\caption[SPECIFIC TO LOF]{diagrams giving rise\label{Diagrams}
to the operators~(\ref{operators}a,b,c)
respectively.}
\end{center}\end{figure}

\section{Yukawa couplings generated by heavy particle exchanges}
{}From the previous section, we are left with two problems:
\begin{itemize}
\item[1.] the need to suppress the operator $15~(\Sigma\bar{H})\bar{6}$
which would tend to $V_{cb}$ close to unity;
\item[2.] the difficulty of distinguishing the masses of the first two
generations.
\end{itemize}
Here we show how both problems can be solved, still without appealing to
any flavour symmetry, by assuming that all the non renormalizable
operators be generated by the exchanges of appropriate heavy particles.

At dimension~5, we only need the operator $20~\Sigma H~15_i$
(irrespective of the possible contractions of the group indices).
This operator can uniquely be generated, as shown in
figure~\ref{Diagrams}a, by the exchange of a~$70_{\rm H}$.
At dimension~6, the operator $20\,\bar{H}\Sigma\bar{H}\bar{6}_i$
generates the masses for the $b$ and the $\tau$.
The simplest possibility, which involves one heavy exchange only, is
shown in figure~\ref{Diagrams}b.
Notice, however that, by cutting this diagram, one has the operator
$(15_{\rm H} \Sigma)\bar{H}\bar{6}$, with the heavy~15 denoted by
$15_{\rm H}$.
In turn, if it were possible to replace~$15_{\rm H}$ by~$15_i$, this
would be a disaster.
Other than matter parity, we are therefore forced to introduce another
${\cal Z}_2$ symmetry, hereafter called $\cal D$, under which
$15_{\rm H}$ (and $\overline{15}_{\rm H}$) change sign, whereas~$15_i$
stay invariant.

Let us now discuss the operators connecting $15_i$ with $\bar{6}_j$.
By restricting ourselves to the exchanges of heavy particles with lower
or upper indices only, and at the same time consistent with $\cal D$-parity,
the only possibility is the one shown in figure~\ref{Diagrams}c,
generating the operator
\begin{equation}
15_i(\Sigma\bar{H})(\Sigma\bar{6}_j)+(15~\bar{H})\Sigma\Sigma\bar{6}_j
\end{equation}
This is what we need, since $\cal D$ parity can be consistently defined
on all multiplets entering the diagrams of fig.~\ref{Diagrams},
as prescribed in table~2, always in a flavour blind way.
Two SU(6)-singlets are included, to
enlarge the Higgs system, which are assumed to get a vacuum expectation value
at a scale $M$, with $\md{H}<M\le M_{\rm Pl}$.
The Yukawa superpotential is taken to be the most general set of
trilinear interactions among the supermultiplets in table~2
compatible with the gauge symmetry, matter parity and $\cal D$-parity.
The heavy supermultiplets acquire their mass at a scale $M$ from the
vacuum expectation values of the singlets $S_1,S_2$.
After integrating out the heavy supermultiplets, other than the
renormalizable superpotential given in eq.~(\ref{eq:W3}), one obtains the
required non renormalizable interactions, as described from the diagrams of
figures~\ref{Diagrams}.
As a matter of fact, from the diagram of fig.~\ref{FalseMass}, one also
obtain the dimension-5 operator $(15_i\bar{H})\Sigma\bar{6}_j$.
Such an operator, however, is irrelevant for the masses of the light
particles, since, in conjunction with the renormalizable term
$15_i\bar{H}\bar{6}'_j$, it only slightly redefines the composition of
the heavy particles.
We shall see in the next section how this construction of the
non renormalizable interactions introduces a distinction between the first
two generations.

\begin{table}[t]
\begin{center}\begin{tabular}{c|l}
$+$ & $S_1,H,\Sigma;\, 20,\bar{6}_i,15_i;~6_{\rm H},
21_{\rm H},\overline{21}_{\rm H},
70_{\rm H},\overline{70}_{\rm H}$\\ \hline
$-$ & $S_2,\bar{H};\,\bar{6}';~\bar{6}_{\rm H},
15_{\rm H},\overline{15}_{\rm H}^{^{}}$
\end{tabular}
\caption{$\cal D$-parity of the various multiplets.}
\end{center}\end{table}
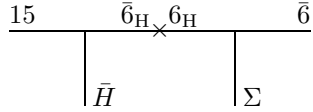
\begin{figure}\setlength{\unitlength}{1cm}
\begin{center}\begin{picture}(6,2)(0,-1)
\put(1,0){\line(1,0){4}}
\multiput(2,0)(2,0){2}{\line(0,-1){1}}
\put(1,0.1){\makebox(0,0)[bl]{15}}
\put(3,0.1){\makebox(0,0)[b]{$\bar{6}_{\rm H}~~6_{\rm H}$}}
\put(5,0.1){\makebox(0,0)[br]{$\bar{6}$}}
\put(2.1,-1){\makebox(0,0)[bl]{$\bar{H}$}}
\put(4.1,-1){\makebox(0,0)[bl]{$\Sigma$}}
\put(3,0){\makebox(0,0){$\times$}}
\end{picture}\caption{the diagram
giving rise to an irrelevant dimension-5 operator.\label{FalseMass}}
\end{center}\end{figure}

\section{A specific ``texture''}
We now show how the specific model defined in section~5 gives rise to
an interesting quasi-realistic texture for fermion masses.
To this purpose, as in section~4, it is useful to redefine,
without loss of generality, appropriate linear combinations of fields
which are not distinguished by any symmetry, gauged or discrete.
With reference to the appropriate terms in the superpotential,
and without writing explicitly the dimensionless couplings,
we do that according to the following progression:
\begin{itemize}
\item[1.] $6_{\rm H} S_2 (\bar{6}_{\rm H} + \bar{6}'_i)\to
6_{\rm H} S_2\,\bar{6}_{\rm H}$ (defines $\bar{6}_{\rm H}$);
\item[2.] $20~H~15_i\to20~H~15_3$ (defines $15_3$);
\item[3.] $15_i\,\Sigma\,\overline{21}_{\rm H}\to
15_{3,2}\,\Sigma\,\overline{21}_{\rm H}$ (defines $15_2$);
\item[4.] $\bar{6}_i\,\bar{H}~15_{\rm H}\to
\bar{6}_3\bar{H}~15_{\rm H}$ (defines $\bar{6}_3$);
\item[5.] $\bar{6}_i\Sigma\,6_{\rm H}\to
\bar{6}_{3,2}\Sigma\,6_{\rm H}$ (defines $\bar{6}_2$).
\end{itemize}
If we know integrate out the heavy vector-like multiplets, we are left
with the renormalizable superpotential
of eq.~(\ref{eq:W3}) and with the following (relevant) non-renormalizable
terms
\begin{eqnsystem}{operators}
&&\lefteqn{\frac{\gamma_i^{(2)}}{M}(20~\Sigma)H~15_i}
\hspace{11em} i=1,2,3\\
&&\lefteqn{\frac{\sigma^{(2)}}{M^2}20~\bar{H}\Sigma\bar{H}~\bar{6}_3} \\
&&\lefteqn{\frac{\sigma_{ab}^{(4)}}{M}15_a(\Sigma\bar{H})(\Sigma\bar{6}_b)}
\hspace{11em} a,b=2,3;\qquad\sigma_{ab}^{(4)}=\delta_a\tau_b
\end{eqnsystem}
respectively associated with the diagrams shown in figure~\ref{Diagrams}a,b,c.
Taking into account that the tree level potential makes heavy the
full~$15_3$, the leading
terms of the $u$, $d$ and $e$ mass matrices are therefore
\begin{eqnsystem}{sys:ude} &&\bordermatrix{& u^c_1&u^c_2 & u^c\cr
Q_1 &0&0&\gamma_1^{(2)}\varepsilon_H\cr
Q_2 &0&0&\gamma_2^{(2)}\varepsilon_H\cr
Q   &\gamma_1^{(2)}\varepsilon_H&\gamma_2^{(2)}\varepsilon_H&
\lambda^{(1)}\cr}\cdot v\sin\beta\\
&&
\bordermatrix{& d^c_1&d^c_2 & d^c\cr
Q_1 &0&0&0&  \cr
Q_2 &0&
  \delta_2\tau_2\varepsilon_\Sigma \varepsilon_H&
  \delta_2\tau_3\varepsilon_\Sigma \varepsilon_H\cr
Q   &0&0&\sigma^{(2)}\varepsilon_H^2}\cdot v\cos\beta \\
&&
\bordermatrix{& L_1&L_2 & L\cr
e_1^c &0&0&0&  \cr
e_2^c &0&
-2\delta_2\tau_2\varepsilon_\Sigma \varepsilon_H&
-2\delta_2\tau_3\varepsilon_\Sigma \varepsilon_H\cr
e^c   &0&0&\sigma^{(2)}\varepsilon_H^2}\cdot v\cos\beta
\end{eqnsystem}
where $\varepsilon_H=\md{H}/M$ and $\varepsilon_\Sigma=\md{\Sigma}/M$.

These mass matrices give rise to a picture for the heavy fermion masses
($t,b,c,\tau$) which is unchanged with respect to section~4.
They actually also constitute an interesting ``texture''.
A simple phase analysis shows, in fact, that the mass matrices~(\ref{sys:ude})
have a single irremovable phase and they involve six other real parameters.
In term of these seven parameters one gets ten flavour observables and
a massless first generation.
Of course, it is also especially important that the size of the various mass
terms is correctly determined by two small parameters only,
$\varepsilon_H\approx10^{-1}$ and $\varepsilon_\Sigma\approx10^{-2}$.

In particular, one obtains the $s,\mu$ Yukawa couplings and the CKM
matrix at the unification scale as
\begin{equation}
\lambda_s\approx\delta_2\tau_2\varepsilon_\Sigma\varepsilon_H,
\qquad\frac{\lambda_\mu}{\lambda_s}\approx 2
\end{equation}
\begin{equation}\label{eq:VCKM}
V_{\rm CKM}\approx\pmatrix{
 1  &s&ss_d\cr
-s  &1&s_ue^{i\varphi}+s_d\cr
ss_u&-(e^{i\varphi}s_d+s_u)&e^{i\varphi}\cr}
\end{equation}
where
\begin{equation}\label{eq:ssusd}
s=\frac{|\gamma_1^{(2)}|}{\sqrt{\gamma_1^{(2)2}+\gamma_2^{(2)2}}},
\qquad
s_d=\varepsilon_\Sigma \varepsilon_H\left|
\frac{\delta_2\tau_3}{\sigma^{(2)} } \right|,
\qquad
s_u=\varepsilon_H\left|\frac{\gamma_2^{(2)}}{\lambda^{(1)}}\right|=
\sqrt{\frac{\lambda_c}{\lambda_t}}
\end{equation}
After the appropriate rescalings to low energy of the various
parameters~\cite{RGE}, one finds a remarkable approximation to a realistic
description of masses and mixings, obtained without appeal to any flavour
symmetry.
Notice in particular the factor of 2 between the muon and the strange quark
masses at unification, characteristic of SU(6), as well as the fact that the
Cabibbo angle, $\theta_{\rm C}$, is naturally of order~1 and the only one to
be unsuppressed. This is perhaps why $\theta_{\rm C}$ is the largest among
the mixing angles.

The masslessness of the first generation is an exact consequence of the
model specified in section~5.
In the
down-lepton sector, the three $\bar{6}_i$ fields are only coupled,
in the superpotential, to two heavy states $6_{\rm H}$ and $15_{\rm H}$,
whereas, in the up sector, one of the $15_i$ fields and the $21_{\rm H}$
are only coupled to the $\overline{21}_{\rm H}$, leading to a massless
linear combination.
The first generation masses, or at least those of the electron and the
down quark, since a massless up quark may be welcome in view of the
strong CP problem, will have to come from a suitable extension of the model.

Let us in fact work under the assumption that the dominant contribution to
first generation masses will come from the~$1,1$ entries in the Yukawa
matrices. Although not guaranteed, within the
democratic approach to flavour, the best guess is that the first generation
masses arise from some higher dimension operator, which is flavour blind and
which then puts a comparable entry everywhere in the mass
matrices~(\ref{sys:ude}).
These entries must then be very small and the dominant contribution to first
generation masses will come from the~$1,1$ entries alone.
In this case, no other diagonalization has to be done to obtain the
quark physical basis.

Under this assumption, the CKM matrix~(\ref{eq:VCKM}),
via eq.~(\ref{eq:ssusd}), leads to an interesting constraint between the
top quark mass and the physical mixing parameters.
In terms of the RG correction factors $y$ and $\eta_c$, defined in
section~4, one obtains
\begin{equation}
A\lambda^2|1-\rho-i\eta|=\sqrt{\frac{ym_c}{\eta_{\rm c}m_t}}
\end{equation}
where $A,\lambda,\rho,\eta$ are the usual Wolfenstein parameters for the
CKM matrix.
Using $y=0.70\pm0.05$, $\eta_c=1.87\div2.30$, with an uncertainty
dominated by the error on $\alpha_3$, and the experimental values
\begin{eqnarray}
\lambda=0.22,\qquad A=0.86\pm0.10\qquad
(|V_{cb}|=0.042\pm0.005),
\end{eqnarray}
this relation can be viewed as a constraint on the lenght $|1-\rho-i\eta|$
of the side of the usual unitarity triangle proportional to $V_{td}$
\begin{equation}
|1-\rho-i\eta|=(1.2\pm0.15)\,\sqrt{\frac{170~{\rm GeV}}{m_t}}
\end{equation}
Such a constraint is perfectly consistent with the present knowledge on
the CKM matrix for a top mass value as in eq.~(\ref{eq:mt})~\cite{Ali}.
It indicates a negative value of $\rho$ ($-0.3\,
\raisebox{-.4ex}{\rlap{$\sim$}} \raisebox{.4ex}{$<$}\,\rho\,
\raisebox{-.4ex}{\rlap{$\sim$}} \raisebox{.4ex}{$<$}0$),
$\sin2\alpha$ close to~1,
$\sin2\beta$ close to~0.5 and a relatively small mixing parameter
in the $B_s$-$\bar{B}_s$ system
$$x_s=(9.8\pm2.5)\frac{m_t}{170\,{\rm GeV}}.$$

To conclude this section we comment on the neutrino masses.
In the multiplets corresponding to one generation,
$\bar{6}\oplus\bar{6}'\oplus15$, there are~5 neutrino states,
3 SU(2)-doublets and~2 SU(2)-singlets.
In the mass matrix, such neutrinos mix among each other and with
neutrinos in the heavy $6,15,21,70$ representations.
It is however readily seen from the general superpotential
and from the absence of neutrinos in the~20, that the neutrino mass
matrix conserves a neutrino number, with opposite charges for the neutrinos
in the barred and unbarred representations.
As a consequence, one is left with~3 massless neutrinos per generation.
Furthermore, by simple inspection one sees that, a part from irrelevant
mixings, only one of these neutrinos (per generation) is a SU(2)-doublet,
necessarily coupled to the light charged leptons.
This situation persists to all order of perturbation theory
until supersymmetry is unbroken.
After supersymmetry breaking, characterized by the scale $m$,
radiative corrections may give rise to
Dirac mass terms involving the SU(2)-doublet neutrinos at most of order
$mv/M$. Although small, since $M
\raisebox{-.4ex}{\rlap{$\sim$}} \raisebox{.4ex}{$>$}10^{17}\,{\rm GeV}$, these
masses could be of phenomenological interest for the solar neutrino physics.

\section{Proton decay}
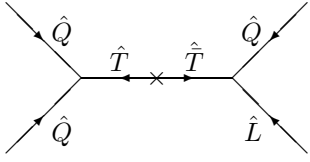
\begin{figure}\setlength{\unitlength}{1mm}
\begin{center}\begin{picture}(40,20)
\put(0,0){\vector(1,1){5}}   \put(5,5){\line(1,1){5}}
\put(0,20){\vector(1,-1){5}}\put(5,15){\line(1,-1){5}}
\put(10,10){\line(1,0){20}}
\put(40,20){\vector(-1,-1){5}} \put(35,15){\line(-1,-1){5}}
\put(40,0){\vector(-1,1){5}}\put(35,5){\line(-1,1){5}}
\put(15,10){\vector(-1,0){0}}\put(15,11){\makebox(0,0)[b]{$\hat T$}}
\put(25,10){\vector(1,0){0}}\put(25,11){\makebox(0,0)[b]{$\hat{\bar T}$}}
\put(6,15){\makebox(0,0)[bl]{$\hat Q$}}
\put(6,5){\makebox(0,0)[tl]{$\hat Q$}}
\put(34,15){\makebox(0,0)[br]{$\hat Q$}}
\put(34,5){\makebox(0,0)[tr]{$\hat L$}}
\put(20,10){\makebox(0,0){$\times$}}
\end{picture}
\caption{the dominant diagram for proton decay in superfield notation.
\label{pDecDiagram}}
\end{center}\end{figure}

Quite in general, the  dominant diagram for proton decay is shown in
figure~\ref{pDecDiagram}
in superfield notation, where $\hat T,\hat {\bar T}$ is the heavy triplet
with the $\SU(3)\otimes\SU(2)\otimes{\rm U}(1)$ quantum numbers of a down
quark~\cite{pDec}. The resulting operator $(\hat Q^a\hat Q^b)(\hat Q^c\hat
L)\varepsilon_{abc}$, where $a,b,c$ are SU(3) indices, leads, after
supersymmetry breaking, to the decay $p\to K\bar\nu$ with a rate that can be
very close to, or even exceed, the present experimental bound~\cite{pDecExp}.
The actual value of this rate depends, other than on the unification scale,
on the masses of some of the superpartners at the Fermi scale and on the
couplings of the heavy triplet to quark and lepton supermultiplets, as shown
in figure~\ref{pDecDiagram}.

As it is the case for the SU(2)  doublets $h_H, \bar{h}_{\bar{H}}$ and
$h_\Sigma, \bar{h}_\Sigma$, in the SU(6) model there are two pairs of
triplets, $T_H,\bar{T}_{\bar H}$ and $T_\Sigma,\bar{T}_\Sigma$.
However, at
variance with the doublets, the triplets appear as Goldstone bosons only in
the breaking of one of the SU(6) factors
$$\SU(6)\stackrel{\md{H}}{\longrightarrow}\SU(5),$$
but not in the other
$$\SU(6)\stackrel{\md{\Sigma}}{\longrightarrow}
\SU(4) \otimes\SU(2)\otimes{\rm U}(1).$$
As a consequence, the triplet $T_H, \bar{T}_{\bar H}$ is eaten by
the SU(6) gauginos, whereas only $T_\Sigma,\bar{T}_\Sigma$ remains as a heavy
state capable of mediating the process of figure~\ref{pDecDiagram}.
Analytically, the difference between $h$'s and $T$'s is clearly seen in the
relevant mass matrices involving also the doublet $\lambda_2,\bar\lambda_2$
and the triplet $\lambda_3,\bar\lambda_3$ gauginos:
$$\bordermatrix{&\lambda_2&h_H&h_\Sigma\cr
\bar\lambda_2&0&\frac{i}{\sqrt{2}}g\md{H}&\frac{3i}{\sqrt{2}}g\md{\Sigma}\cr
\bar{h}_{\bar H}&-\frac{i}{\sqrt{2}}g\md{H}&0&0\cr
\bar{h}_\Sigma&-\frac{3i}{\sqrt{2}}g\md{\Sigma}&0&0},\qquad
\bordermatrix{&\lambda_3&T_H&T_\Sigma\cr
\bar\lambda_3&0&\frac{i}{\sqrt{2}}g\md{H}&0\cr
\bar{T}_{\bar H}&-\frac{i}{\sqrt{2}}g\md{H}&0&0\cr
\bar{T}_\Sigma&0&0&\lambda\md{\Sigma}}.$$
The masses  coming from the $d$-terms are proportional to the gauge coupling
$g$, whereas the mass term  $\lambda\md{\Sigma}$ comes from the
superpotential. While the doublets from $H,\bar H$ and $\Sigma$ are mixed by
the mass matrix, the triplets are not.
This property has the remarkable consequence that, whenever the light Higgs
doublet that provides the relevant mass term comes from $H, \bar H$, the
corresponding triplet coupling is ineffective to the proton decay. An example
is provided by the coupling proportional to $\gamma^{(5)}$ in
eq.~(\ref{eq:W4}).
It is possible to conceive explicit flavour structures of
the Yukawa couplings to the light generation which are at the same time
realistic and lead to a suppression of the $p\to K\bar\nu$ amplitude by a
factor $\md{\Sigma}/\md{H}\approx 0.1$ or by its square.
Such a suppression might be important in view of the existing estimates of
the proton decay lifetime in the minimal SU(5) model~\cite{pDecRecent}.

\section{Conclusions}
In this paper we wanted to analyse how far one could go in accounting
for the observed pattern of fermion masses and mixings in a
grand unified theory without any flavour symmetry.
We have studied the problem in an interesting model for the
doublet-triplet splitting.
To our surprise we have found an elegant understanding of the
heaviness of the top quark.
Instrumental to this result is the fact that the top comes out
belonging to an irreducible representation of the gauge group
not isomorphic to those ones containing the other quarks of charge $2/3$.

In own framework the splitting of the heavy top from the light fermions
results from simply writing down the most general gauge invariant
renormalizable Lagrangian.
Extending this to non-renormalizable terms does not immediately lead to
a satisfactory generation of the lighter fermion masses.
However, by including a flavour-blind discrete symmetry and requiring that all
higher dimensional operators be mediated by the exchange of appropriate heavy
multiplets, it is possible to give an approximate description of all
masses and mixings in terms of the hierarchy $M>\md{H}>\md{\Sigma}$.
This leads to several interesting features.
One finds $\lambda_b/\lambda_t\approx\varepsilon_H^2$ and
$\lambda_b/\lambda_\tau=1$ which is successful provided $\lambda_t$ is
very close to its infrared fixed point. As a consequence, $m_t$ and $m_h$
are correlated in a specific way.
The light quark mass hierarchies are understood as
$m_b/m_t\approx m_c/m_t\approx\varepsilon_H^2$ and
$m_s/m_t\approx\varepsilon_H\varepsilon_\Sigma$.
In addition to the GUT relation $m_b/m_\tau=1$,the SU(6) theory
predicts $m_s/m_\mu=1/2$.
A characteristic ``texture'' leads to a Kobayashi-Maskawa mixing matrix
in which the magnitude of $V_{td}$ can be predicted at the $15\,$\% level
of accuracy.
This leads to a prediction for $x_s$ and the CP violation angles
$\sin2\alpha$ and $\sin2\beta$ in $\beta$ decays.

Finally, we have pointed out a possible reason for the suppression of the
$p\to K\bar\nu$ amplitude, intimately related to the mechanism of the
doublet-triplet splitting.
Needless to say, the next step is to find a source for the first generation
masses.

\section*{Acknowledgements}
Z.B.\ thanks Alexei Anselm for useful comments.

\nonfrenchspacing
\end{document}